\documentclass[11pt]{article}
\usepackage{amsmath,amssymb,color,graphics,epsfig,cite}
\usepackage{amsfonts}
\newcommand{\hoch}[1]{$\, ^{#1}$}

\newcommand{\be}{\begin{equation}}
\newcommand{\ee}{\end{equation}}
\newcommand{\bea}{\setlength\arraycolsep{2pt} \begin{eqnarray}}
\newcommand{\eea}{\end{eqnarray}}
\newcommand{\nn}{\nonumber}

\def\ft#1#2{{\textstyle{\frac{\scriptstyle #1}{\scriptstyle #2} } }}
\def\fft#1#2{{\frac{#1}{#2}}}

\def\0{{\sst{(0)}}}
\def\1{{\sst{(1)}}}
\def\2{{\sst{(2)}}}
\def\3{{\sst{(3)}}}
\def\4{{\sst{(4)}}}
\def\5{{\sst{(5)}}}
\def\6{{\sst{(6)}}}
\def\7{{\sst{(7)}}}
\def\8{{\sst{(8)}}}
\def\sst#1{{\scriptscriptstyle #1}}

\begin{document}

\begin{center}

{\Large {\bf Black $p$-brane Thermodynamics \\ without Constructing Solutions}}

\vspace{20pt}

Bing-Yang Han\hoch{1}, H. L\"u\hoch{1,2}

\vspace{10pt}

{\it \hoch{1}Center for Joint Quantum Studies, Department of Physics,\\
			School of Science, Tianjin University, Tianjin 300350, China }

\medskip

{\it \hoch{2}The International Joint Institute of Tianjin University, Fuzhou,\\ Tianjin University, Tianjin 300072, China}

\vspace{40pt}

\underline{ABSTRACT}
\end{center}

\vspace{10pt}

This paper generalizes the method used in the previous article \cite{Yang:2025rud} to black $p$-brane thermodynamics in arbitrary dimensions containing black holes and strings as special cases: thermodynamic quantities can be derived without constructing the corresponding black $p$-brane solutions. We further extend the discussion to black holes or $p$-branes involving a general scalar coset.

\vfill {\footnotesize han\_bingyang@tju.edu.cn\ \ \ mrhonglu@gmail.com}

\thispagestyle{empty}
\pagebreak
%\voffset=-40pt
%\setcounter{page}{1}

%\tableofcontents
%\addtocontents{toc}{\protect\setcounter{tocdepth}{2}}

%\newpage

\section{Introduction}

In strings or M-theory, the fundamental objects are no longer particles, but extended objects generally referred to as (super) $p$-branes \cite{Achucarro:1987nc}. They manifest themselves as soliton-like solutions in supergravities \cite{Duff:1994an}, which are the leading low-energy effective theories. A $p$-brane soliton splits the $D$-dimensional spacetime into the world-volume of dimensions $d=p+1$ and transverse space of dimensions $(D-d)$. The initial constructions focused on supersymmetric $p$-branes, including string soliton \cite{Dabholkar:1990yf}, M-branes \cite{Gueven:1992hh} and D3-branes \cite{Duff:1991pea}. There are a number of classifications of the $p$-branes \cite{Lu:1995cs} and their intersections \cite{Papadopoulos:1996uq,Klebanov:1996mh} in the literature, e.g.~\cite{Behrndt:1996pm,Lu:1997hb}.

The black $p$-brane solutions with non-vanishing temperature were later constructed \cite{Duff:1996hp,Cvetic:1996gq}, which lead to the study of black $p$-brane thermodynamics \cite{Muto:1996xv}, generalizing the topic of black hole thermodynamics to extended black objects. On the other hand, in the context of supergravities, many black holes in lower dimensions can be viewed as intersecting fundamental $p$-branes in strings or M-theory \cite{Papadopoulos:1996uq,Klebanov:1996mh}, based on which, the first example of microscopic counting of extremal or near extremal black hole entropy was achieved \cite{Strominger:1996sh}. Thus a full study of black $p$-brane thermodynamics can shed new light on the lower-dimensional black holes.

However, the exact solutions of black $p$-branes constructed in \cite{Duff:1996hp, Cvetic:1996gq} are only small samples of possible allowed black $p$-branes that exist in supergravities. The majority of black $p$-branes either do not have exact solutions or such solutions are not yet known. To study their thermodynamics by constructing numerical solutions is a tedious task and requires a strong motivation. It was recently shown that black hole thermodynamics for large classes of generalized Einstein-Maxwell-Dilaton (GEMD) theories can be completely derived without constructing the corresponding black hole solutions \cite{Lu:2025eub,Yang:2025rud}. The validity of the thermodynamic quantities were further confirmed  \cite{Lu:2026lkv} by exact Toda black holes \cite{Lu:2026lkv,Ivashchuk:2013jja}. The GEMD theories can either be directly embedded in or inspired by supergravities. This leads to a natural question: can we develop the same mechanism for black $p$-brane thermodyanmics for the more fundamental extended black $p$-branes without having to construct the solutions? In this paper, we investigate this and provide explicit formulae to derive all the thermodynamic quantities of the black $p$-branes that are spherically-symmetric and asymptotic to Minkowski spacetimes. The generalization is not as straightforward as it might appear; the subtlety lies in the long-range force of two identical black $p$-branes. Fortunately, this issue was studied and resolved in \cite{Heidenreich:2020upe}.

The paper is organized as follows. In Section 2, we consider Lagrangians that are direct generalizations of GEMD theories where all the Maxwell fields are replaced by the high-form fields that are responsible to charge the extended objects. We propose a set of equations on the thermodynamic quantities, which can be used to derive all the thermodynamic quantities without constructing the solutions. We then solve these equations in Section 3 for special choices of dilaton couplings and show that they precisely reproduce the corresponding thermodynamics in literature from the explicit solutions. In Section 4, we consider the more general theory where the high-form fields are coupled to a scalar coset and propose the equations on the corresponding black $p$-brane thermodynamic quantities. We then use an $SL(2,\mathbb R)/SO(2)$ example to illustrate how to obtain the black $p$-brane thermodynamics without using the black $p$-brane solution. We conclude the paper in Section 5.

\section{General formalism}

\subsection{Theory and ansatz}

We are concerned with supergravities that are low-energy effective theories of strings or M-theory, in a toroidal compactification keeping only the zero modes. The bosonic field content typically consists of a scalar coset ${\cal G}/K$, together with a set of form fields of different ranks. For each given rank $(p+2)$, the form fields belong to a specific representation of ${\cal G}$. The scalar coset involves both dilatonic and axionic scalars. In this section, we shall focus on the truncation where the axionic scalars are set to zero. The discussion of the more general cosets is relegated to Section 4. The Lagrangian takes the form
\begin{equation}
	\mathcal{L}=\sqrt{-g}\left[R-\frac{1}{2}(\partial\vec{\phi})^2 -\frac{1}{2\cdot(p+2)!}\sum_{\alpha=1}^{N}\mathrm{e}^{\vec{a}_\alpha\cdot\vec{\phi}}
F_{\alpha}^2\right],\qquad F_\alpha=dA_\alpha\,.\label{generallag1}
\end{equation}
Here, we assume that the total number of scalar fields is less than $N$, since if we have more scalars, the scalar vector $(\vec \phi)_\perp$ that is perpendicular to all $\vec a_\alpha$ can be truncated out. The Lagrangian \eqref{generallag1} can be embedded in supergravity only for some specific choice of $\vec a_\alpha$ and $N$. (See, e.g.~\cite{Lu:1997hb} for classification.) Thus, for generic $\vec a_\alpha$ and $N$, the Lagrangian describes a string-inspired theory.

In supergravities, a $k$-form field strength supports an electrically-charged $(k-2)$-brane or a magnetically-charged $(D-k-2)$-brane. For example, in $D=11$ dimensions, the 4-form field strength is responsible for the electric M2-brane or the magnetic M5-brane. We named our form fields in the above as the $(p+2)$-form since we should focus on the discussion of electrically-charged $p$-branes, as their magnetic duals follow straightforwardly.  In this paper, we consider spherically-symmetric black $p$-branes where the world-volume is Minkowski spacetime with a blackening factor. The general non-extremal multi-charged electric black $p$-brane ansatz takes the form \cite{Duff:1996hp}
\bea
ds^2 &=& e^{2A(r)} \Big(-h(r)(dt)^2 + (dx_1)^2 + \cdots (dx_p)^2\Big) + e^{2B(r)}
\Big(\fft{dr^2}{h(r)} + r^2 d\Omega_{\tilde p+2}^2\Big),\nn\\
A_\alpha &=& \Phi_\alpha(r) dt\wedge dx_1\wedge \cdots\wedge dx_p\,,\quad \alpha=1,2,\ldots, N\,,\qquad
\vec \phi = \vec \phi(r)\,.\label{pansatz}
\eea
Note that we introduced a parameter $\tilde p$, defined by
\be
p+\tilde p=D-4\,.
\ee
The parameter $\tilde p$ describes the spatial dimension of the world-volume of the magnetic $\tilde p$-brane. Although we do not consider magnetic branes in this paper, we find that the $\tilde p$ parameter is useful to have.

When all the matter fields are turned off, the solution becomes Ricci flat, describing a neutral black $p$-brane, with
\be
h=1 - \fft{2\nu}{r^{\tilde p+1}}\,.
\ee
The temperature and mass, entropy densities are
\be
T=\fft{\tilde p+1}{4\pi r_+}\,,\qquad
M=\fft{(\tilde p+2) \Omega_{\tilde p+2}}{8\pi}\, \nu\,,\qquad
S=\ft14 \Omega_{\tilde p+2}\, r_+^{\tilde p+2}\,,
\ee
which satisfy the first law and Smarr relation, i.e.
\be
dM=T dS\,,\qquad M = \fft{\tilde p+2}{\tilde p+1} T S\,.
\ee
When the $p$-brane volume is compact, we assume that its volume is unity; if it is non-compact, the extensive quantities such as mass, entropy and charges are all of the density type, as quantities per unit volume. (Note that for non-compact world-volume, such a neutral black string or $p$-brane is unstable \cite{Gregory:1993vy}.)

For the general charged black $p$-brane, we can keep the function $h$ unchanged as the blackening factor while the functions $(A,B,\Phi_\alpha, \vec \phi)$ are solved as functions of the radial coordinate $r$. Since all the fields are massless, we have a total of $(N+n+1)$ hairy parameters, given by
\be
g_{tt}= -1 + \fft{2m}{r^{\tilde p+1}} +\cdots\,,\qquad \vec \phi = \fft{4\vec \sigma}{r^{\tilde p+1}} + \cdots \,,\qquad \Phi_\alpha = \fft{4q_\alpha}{(\tilde p+1) r^{\tilde p+1}} + \cdots\,.\label{falloffs}
\ee
In this paper, we set $\vec \phi(\infty)=0$. We therefore have the usual ADM mass $M$ \cite{Lu:1993vt}. In addition, there are $N$ electric charges and $n$ scalar charges, defined by
\be
Q_\alpha =\fft{\Omega_{\tilde p+2}}{4\pi} q_\alpha \,,\qquad
\vec \Sigma = \fft{(\tilde p+1) \Omega_{\tilde p+2} }{4\pi} \,\vec \sigma\,.
\ee
For latter purpose, we further introduce two parameters $(\delta, \tilde \delta)$, defined by
\be
\delta = \fft{2(p+1)(\tilde p+1)}{D-2}\,,\qquad
\tilde\delta = \fft{2(\tilde p +1)}{\tilde p+2}\,.
\ee
The first term was expressed as $2d\tilde d/(D-2)$ in \cite{Duff:1994an}, where $d=p+1$ and $\tilde d=\tilde p+1$. It was shown in literature that the black $p$-branes satisfy \cite{Lu:1995cs,Duff:1996hp}
\be
(p+1) A + (\tilde p+1) B=0\,,
\ee
which implies that the product $T S$ is determined by the function $h(r)$ only, namely
$T S= \fft{h'(r_+)}{4\pi}$. We thus follow \cite{Lu:2025eub,Yang:2025rud,Lu:2026lkv} and introduce a parameter $\mu$, given by
\be
\mu =\fft{(\tilde p+2)\Omega_{\tilde p+2}}{8\pi} \nu\,\qquad \longrightarrow\qquad
T S = \ft12 \tilde \delta \mu\,.\label{TSmu}
\ee
We refer to this as the ``asymptotic-horizon relation,'' which we derived without having to solve all the equations. For the neutral black $p$-brane, we simply have $M=\mu$. It is understood that although we shall show that the charged black $p$-brane thermodynamics can be derived without solutions, the neutral black $p$-brane solution is known and used.

\subsection{Derivation of thermodynamics}

For black holes ($p=0$), it was shown that their thermodynamics can be derived without having to construct the solutions \cite{Lu:2025eub,Yang:2025rud,Lu:2026lkv}. Here we extend the discussion to include the general black $p$-branes of the type \eqref{pansatz}. One key advance is to find suitable variables to describe the black $p$-brane. The general solution contains $N+n+1$ parameters, $(M, Q_\alpha, \Sigma_i)$. We expect that the weak version \cite{Lu:2025eub} of no-scalar-hair theorem applies so that the scalar charges $\Sigma_i$'s are functions of mass $M$ and $N$ electric $(p+2)$-form charges $Q_\alpha$'s. All the thermodynamic quantities are then functions of $(M,Q_\alpha)$. However, this is not the most convenient way of expressing the thermodynamic quantities, as they may not have a close analytic form even for some exact solutions. This issue was resolved for black holes by introducing a parameter $\mu$ that describes the strength of long-range force between two identical such black holes. This concept can be generalized to include black $p$-branes, but subtleties emerge.

\textbf{Long-range force law.} It turns out that the long-range force between two identical black $p$-branes also involves the tension force; it is given by \cite{Heidenreich:2020upe}
\begin{align}
	\lim_{r\to\infty}r^{\tilde{p}+2}\text{Force}
	=-2\left[M^2+p\mathcal{T}^2-\frac{(M+p \mathcal{T})^2}{D-2}\right]-4\vec \Sigma\cdot \vec \Sigma+4\sum_{\alpha=1}^N Q_\alpha^2\,,
\label{lrfl}
\end{align}
where ${\cal T}$ is the ADM tension. For our black $p$-brane ansatz \eqref{pansatz}, the tension is also the free energy, namely ${\cal T}=M-TS$ \cite{Heidenreich:2020upe}. While this quantity ${\cal T}$ can be defined for black holes $(p=0)$, its contribution to the long-range force drops out for black holes.

It follows from \eqref{TSmu} that we can now replace $TS$ by the parameter $\mu$. Thus, we have
\be
{\cal T} = M - \ft12 \tilde \delta \mu\,.\label{tensionmu}
\ee
Substituting the leading falloffs of the black $p$-brane ansatz \eqref{pansatz} into the equations of motion, we can easily show that the total strength of the long-range force
\eqref{lrfl} is simply $-\tilde \delta \mu^2$. We therefore obtain the new long-range force law between two identical $p$-branes:
\be
\Big[M + \Big(\fft{\tilde\delta}{\delta}-1\Big)\mu\Big]^2 +
\fft{4}{\delta}\Big[\vec \Sigma \cdot \vec \Sigma - \sum_{\alpha=1}^N Q_\alpha^2\Big] =
\Big(\fft{\tilde\delta}{\delta}\mu\Big)^2\,.\label{lrfl2}
\ee
It is easy to verify that the neutral black $p$-brane satisfies this condition. This equation involves all the nontrivial asymptotic parameters $(M, \mu, Q_\alpha, \vec \Sigma)$. It implies that we can now replace $M$ with $\mu$ as a basic parameter for thermodynamic quantities. This provides a key step towards deriving the complete set of thermodynamic quantities.

It is important to note that constraint \eqref{lrfl2} generalizes \cite{Heidenreich:2020upe}: it is valid for general ansatz \eqref{pansatz}, even when the solution is not a black $p$-brane, but with a naked singularity; these singular $p$-brane solutions arise when the scalar hair parameters $\vec \Sigma$ become independent and the asymptotic-horizon relation \eqref{TSmu} is no longer valid. This is consistent with the concept of long-ranged force, which should not care whether the spacetime is a black $p$-brane or not.

\textbf{Weak no-hair theorem conjecture.} A general spherically-symmetric and static $p$-brane solution contains $n$ independent scalar charge, in addition to the mass and $N$ electric charges. Such a general solution has a naked curvature singularity. For the solution to describe a black $p$-brane that is regular from the horizon to asymptotic infinity, the weak no-hair theorem conjecture \cite{Lu:2025eub} should be assumed.
In other words, the scalar charge vector $\vec \Sigma$ is not independent, but a specific vector function of $(M, Q_\alpha)$. The long-range force condition implies that we can replace $(M,Q_\alpha)$ with $(\mu, Q_\alpha$). Following the same argument of \cite{Lu:2025eub}, we have
\begin{equation}
	\vec \Sigma=\frac{1}{2} \sum_{\alpha=1}^N \vec a_\alpha Q_\alpha \frac{\partial M}{\partial Q_\alpha}+\mu\frac{\partial\vec \Sigma}{\partial\mu}\,,\label{mcr}
\end{equation}
with a boundary condition that $\vec \Sigma=0$ when all the electric charges are turned off. It is worth commenting that the above relation continues to hold when we replace the ADM mass $M$ with the ADM tension ${\cal T}$ given by \eqref{tensionmu}.

\textbf{Homogeneity.} Another piece of clue is that for theories such as \eqref{generallag1}, any thermodynamic quantity $X$ should be homogeneous under the scaling of the basic variables $(\mu, Q_\alpha)$, namely
\begin{equation}
	X(\lambda \mu,\lambda Q_\alpha)=\lambda^\epsilon X(\mu, Q_\alpha),
\end{equation}
where $\epsilon$ is the dimension relative to $(\mu, Q_\alpha)$, which have the same dimension. In this paper, the notation $X(Q_\alpha)$ denotes a function $X$ depending on all the charges $Q_\alpha$, $\alpha=1,2,\ldots, N$. The notation $X_\alpha(Q_\alpha)$, however, denotes a function $X_\alpha$ depending on $Q_\alpha$ only. Note that the scalar charge vector $\vec\Sigma$ has the same dimension as the basic variables, which implies that \eqref{mcr} can be written as
\begin{equation}
\sum_{\alpha=1}^N Q_\alpha \frac{\partial}{\partial Q_\alpha} \Big(\vec \Sigma - \fft12 \vec a_\alpha M\Big)=0\,.\label{mcr1}
\end{equation}
Together with \eqref{lrfl2}, we can determine mass $M$ and scalar charge vector $\vec \Sigma$ as functions of the basic variables $(\mu, Q_\alpha)$.

\textbf{The first law.} Finally, we assume that the first law of black $p$-brane thermodynamics is, after taking $\vec\phi(\infty)=0)$,
\begin{equation}
	dM=T\, dS+\sum_{\alpha=1}^{N}\Phi_\alpha\, dQ_\alpha\,,\label{fl}
\end{equation}
which can be guaranteed by the Wald formalism. It follows from \eqref{TSmu} and the solved $M(\mu, Q_\alpha)$, the temperature, entropy and electric potentials can be obtained, given by
\begin{equation}
	T=\frac{\tilde{\delta}\mu}{2S}\,,\qquad
	\frac{\partial\ln S}{\partial\mu}
	=\frac{2}{\tilde{\delta}\mu}\frac{\partial M}{\partial\mu}\,,\qquad
	\Phi_{\alpha}=\frac{\partial M}{\partial Q_{\alpha}}-\frac{\tilde{\delta}\mu}{2}\frac{\partial\ln S}{\partial Q_{\alpha}}\,,\qquad \alpha=1,2,...,N.\label{fle}
\end{equation}
We now have constructed all the equations needed for deriving thermodynamic quantities.

\textbf{Solving the thermodynamic equations.} The homogeneity condition inspires us to introduce dimensionless parameters $x_\alpha = Q_\alpha/\mu$ and dimensionless functions for the corresponding thermodynamic quantities, namely
\be
M= \mu\Big(f(x_\beta)+1-\frac{\tilde{\delta}}{\delta}\Big),\qquad \vec{\Sigma}=\mu\, \vec{g}(x_\beta)\,,\qquad
S=\mu^{2/\tilde{\delta}}\, s(x_\beta)\,,\qquad \Phi_{\alpha}=\Phi_{\alpha}(x_\beta)\,.
\label{allthermo}
\ee
The equations \eqref{lrfl2} and \eqref{mcr1} become respectively
\be
f^2+\frac{4}{\delta}\left(\vec{g}\cdot\vec{g}-\sum_{\alpha=1}^{N}x_{\alpha}^2\right)
=\left(\frac{\tilde{\delta}}{\delta}\right)^2\,,\qquad 	\sum_{\alpha=1}^{N}x_{\alpha}\frac{\partial}{\partial x_{\alpha}}\left(\vec{g}-\frac{1}{2}\vec{a}_{\alpha}f\right)=0\,,\label{fgeq}
\ee
which determine functions $(f,\vec g)$, and hence ultimately $(M, \vec \Sigma)$.  Once the function $f$ is determined, the dimensionless entropy and electric potentials are given by
\be
\sum_{\alpha=1}^{N}x_{\alpha}\frac{\partial\ln s}{\partial x_{\alpha}}=\frac{2}{\delta}-\frac{2}{\tilde{\delta}}
\left(f-\sum_{\alpha=1}^{N}x_{\alpha}\frac{\partial f}{\partial x_{\alpha}}\right)\,,\qquad \Phi_{\alpha}=\frac{\partial f}{\partial x_{\alpha}}-\frac{\tilde{\delta}}{2}\frac{\partial\ln s}{\partial x_{\alpha}}\,.
\label{sphi}
\ee
The equations \eqref{fgeq} and \eqref{sphi} involve differential equations that lead to arbitrary integration constants, which should be fixed by proper boundary conditions. We shall use the thermodynamics of the neutral black $p$-brane as the boundary conditions, which are consistent with the equations \eqref{fgeq} and \eqref{sphi}. We therefore have the universal boundary conditions
\be
f(0)= \fft{\tilde \delta}{\delta}\,,\qquad s(0)=s_0\equiv \left(\frac{4}{\Omega_{\tilde{p}+2}}\right)^{\frac{1}{\tilde{p}+1}}
\left(\frac{4\pi}{\tilde{p}+2}\right)^{\frac{2}{\tilde{\delta}}}\,,\qquad
\vec g(0)=0\,,\qquad\Phi_\alpha(0)=0\,,\label{boundary}
\ee
for all $\alpha$. With these, the black $p$-brane thermodynamics can be completely determined.

\section{Explicit examples}

\subsection{Single-charge black $p$-brane}

In this subsection, we consider the Lagrangian involving a single form field:
\begin{equation}
	\mathcal{L}=\sqrt{-g}\left[R-\frac{1}{2}(\partial\phi)^2
	-\frac{1}{2\cdot(p+2)!}\mathrm{e}^{a\phi}F_{(p+2)}^2\right],\label{scl1}
\end{equation}
where $F_{(p+2)}=dA_{(p+1)}$ is the field strength of the $(p+1)$-form potential $A_{(p+1)}$.

\subsubsection{Dilatonless black $p$-brane}

We first consider $a=0$, in which case the dilaton decouples, giving rise to dilatonless black $p$-branes. These include the non-extremal M2, M5-branes and D3-branes, etc. It follows from \eqref{fgeq} and \eqref{sphi}, after imposing the necessary boundary conditions in \eqref{boundary}, that
\be
f(x) = \sqrt{\frac{4}{\delta}x^2+\left(\frac{\tilde{\delta}}{\delta}\right)^2}\,,\qquad
s = s_0\left(\frac{1}{2}+\sqrt{\frac{1}{4}+\frac{\delta}{\tilde{\delta}^2}x^2}
\right)^{\frac{2}{\delta}}\,,\qquad \Phi=\frac{4x}{\tilde{\delta}+\sqrt{\tilde{\delta}^2+4\delta x^2}}\,,
\ee
together with $g(x)=0$. All the thermodynamic quantities follow immediately from \eqref{allthermo}. After comparing these results with the known exact solution \cite{Duff:1996hp}, we find that one can identify them by redefining the parameters
\begin{equation}
	k\to\frac{16\pi\mu}{(\tilde{p}+2)\Omega_{\tilde{p}+2}}\,,\qquad \sinh^2\mu\to\sqrt{\frac{1}{4}+\delta\left(\frac{Q}{\tilde{\delta}\mu}\right)^2}-\frac{1}{2}\,,
\end{equation}
where the parameters in the left hand of the identifications are parameters used in \cite{Duff:1996hp}.

\subsubsection{Dilatonic black $p$-brane}

The majority of single-charge black $p$-branes involve dilatons. For $a\ne 0$, the two equations in \eqref{fgeq} can also be solved exactly, given by
\be
g(x)=\frac{a}{2}f(x)+C\,,\qquad f(x)=-\frac{2a}{\Delta}C\pm\sqrt{\frac{4}{\Delta}
\left(x^2+\frac{\tilde{\delta}^2}{4\delta}\right)-\frac{4\delta}{\Delta^2}C^2}\,,
\ee
where $\Delta=a^2 + \delta$. The boundary condition $g(0)=0$ determines the integration constant $C=-a\tilde \delta/(2\delta)$. Consequently, equations in \eqref{sphi} can be solved, giving to
\be s=s_0\left(\frac{1}{2}+\sqrt{\frac{1}{4}+\frac{\Delta}{\tilde{\delta}^2}x^2}
\right)^{\frac{2}{\Delta}}\,,\qquad \Phi=\frac{4x}{\tilde{\delta}+\sqrt{\tilde{\delta}^2+4\Delta x^2}},
\ee
We thus have
\bea
\Sigma &=& -\frac{a\tilde{\delta}}{2\Delta}\mu+\frac{a}{2}\sqrt{\frac{4}{\Delta}Q^2
+\left(\frac{\tilde{\delta}}{\Delta}\mu\right)^2}\,,\qquad M = \left(1-\frac{\tilde{\delta}}{\Delta}\right)\mu+\sqrt{\frac{4}{\Delta}Q^2
+\left(\frac{\tilde{\delta}}{\Delta}\mu\right)^2}\,,\nn\\
S &=& s_0\mu^{\frac{2}{\tilde{\delta}}}\left[\frac{1}{2}+\sqrt{\frac{1}{4}
+\Delta\left(\frac{Q}{\tilde{\delta}\mu}\right)^2}\right]^{\frac{2}{\Delta}}\,,\qquad
\Phi=\frac{4Q}{\tilde{\delta}\mu+\sqrt{(\tilde{\delta}\mu)^2+4\Delta Q^2}}\,.
\eea
Comparing these with the known exact solutions in \cite{Duff:1996hp}, we find that they will coincide after redefining
\begin{equation}
	k\to\frac{16\pi\mu}{(\tilde{p}+2)\Omega_{\tilde{p}+2}}\,,\qquad \sinh^2\mu\to\sqrt{\frac{1}{4}+\Delta\left(\frac{Q}{\tilde{\delta}\mu}\right)^2}
-\frac{1}{2}\,.
\end{equation}

\subsection{Dilatonic multi-charge $p$-branes}

We are now dealing with the general case \eqref{generallag1}, but with an assumption that the number of scalar fields is the same as the number of the form fields. In other words, all the components of the $N$ dilaton vectors $(\vec a_\alpha)^i$ form an $N\times N$ invertible matrix. The equations \eqref{fgeq} and \eqref{sphi} cannot be solved in general and we make a ``non-interacting'' ansatz
\be
f(x_\alpha) = \sum_{\alpha=1}^N f_\alpha (x_\alpha) + f_0\,,\qquad
\vec g = \ft12 \sum_{\alpha}^N \vec a_\alpha f_\alpha (x_\alpha)\,,
\ee
where we impose $f_\alpha(0)=0$ for all $\alpha$ and the universal boundary conditions \eqref{boundary}. Note that $f(x_\alpha)$ is a function of all $x_\alpha$, whilst $f_\alpha(x_\alpha)$ is a function of the specific $x_\alpha$. The consistency of the equation \eqref{fgeq} implies that the dilaton coupling constants must take the form
\be
{\cal A}_{\alpha\beta} \equiv \vec a_\alpha \cdot \vec a_{\beta} =
\Delta' \delta_{\alpha\beta} - \delta\,.\label{adota}
\ee
where $\Delta'$ can be arbitrary real constants. In particular, the $\Delta'=4$ case emerges naturally in supergravities, for which exact solutions were constructed \cite{Duff:1996hp}. The dot product assumption of the dilaton vectors leads to
\begin{equation}
f_\alpha(x_{\alpha})=-\frac{\tilde{\delta}}{\Delta'}+
\sqrt{\frac{4}{\Delta'}x_{\alpha}^2+\left(\frac{\tilde{\delta}}{\Delta'}\right)^2}\,.
\end{equation}
It is then straightforward to obtain the complete set of scalar charges and thermodynamic quantities
\bea
\vec{\Sigma}&=&\sum_{\alpha=1}^{N}\Big[-\frac{\vec{a}_\alpha
\tilde{\delta}}{2\Delta'}\mu+\frac{\vec{a}_\alpha}{2}
\sqrt{\ft{4}{\Delta'}Q_{\alpha}^2+
\left(\ft{\tilde{\delta}}{\Delta'}\mu\right)^2}\Big],\quad M = \left(1-\frac{N\tilde{\delta}}{\Delta'}\right)\mu+\sum_{\alpha=1}^{N}
\sqrt{\ft{4}{\Delta'}Q_{\alpha}^2+\left(\ft{\tilde{\delta}}{\Delta'}\mu\right)^2}\,,\nn\\
%%%%%%%%
S &=& s_0\mu^{\frac{2}{\tilde{\delta}}}\prod_{\alpha=1}^{n}
\left[\frac{1}{2}+\sqrt{\ft{1}{4}+\Delta'\left(\ft{Q_\alpha}
{\tilde{\delta}\mu}\right)^2}\right]^{\frac{2}{\Delta'}}\,,\qquad \Phi_{\alpha}=\frac{4Q_{\alpha}}{\tilde{\delta}\mu+
\sqrt{(\tilde{\delta}\mu)^2+4\Delta'Q_{\alpha}^2}}\,.
\eea
When $\Delta'=4$, the above results reproduce the multi-charged generalization studied in \cite{Duff:1996hp} after performing the reparameterization
\begin{equation}
	k\to\frac{16\pi\mu}{(\tilde{p}+2)\Omega_{\tilde{p}+2}}\,,\qquad \sinh^2\mu_\alpha\to\sqrt{\frac{1}{4}+
\Delta'\left(\frac{Q_\alpha}{\tilde{\delta}\mu}\right)^2}-\frac{1}{2}\,.
\end{equation}
Thus, using the thermodynamic equations of Section 2, we verified that all the thermodynamic quantities previously obtained from black $p$-branes can now be directly derived without these solutions.

\section{Black $p$-brane from coset structures}

\subsection{General formulae}

In Section 2, we have mentioned that the bosonic sector of a supergravity action typically involves a scalar coset structure; the $(p+2)$-form field strengths form a representation. The Lagrangian takes the form
\begin{equation}
	\mathcal{L}=\sqrt{-g}\left(R
	+\frac{1}{4}{\rm tr}(\partial\mathcal{M}\partial\mathcal{M}^{-1})
	-\frac{1}{2\cdot(p+2)!}F^T\mathcal{M}F\right)\,,\label{genlag2}
\end{equation}
where
\begin{equation}
	\mathcal{M}=\mathcal{V}^{\rm T}\mathcal{V}\,,\qquad F=(F_1,F_2,...,F_N)^{\mathrm{T}},
\end{equation}
and each $F_{\alpha}=dA_{\alpha}$ is a field strength of a $(p+1)$-form potential $A_{\alpha}$.
The theory is invariant under some global transformations
\begin{equation}
	\mathcal{M}\to \Lambda\mathcal{M}\Lambda^{\mathrm{T}}\,,\qquad \ F\to (\Lambda^{\mathrm{T}})^{-1}F.
\end{equation}
Note that the symbol ``T'' denotes transpose for orthogonal or special linear groups. For more complicated groups, ``T'' is an involution describing the generalized transpose \cite{Cremmer:1997ct}.

The scalar coset Lagrangian belongs to a class of nonlinear $\sigma$-model, and it can also be expressed as
\begin{equation}
\frac{1}{4}{\rm tr}(\partial\mathcal{M}\partial\mathcal{M}^{-1})=
-\fft12 \sum_{i,j=1}^n G_{jk}(\phi)\partial_{\nu}\phi^j\partial^{\nu}\phi^k\,,
\end{equation}
Here we assume that the dimension of the scalar coset is $n$. A scalar coset typically consists of both dilatonic and axionic scalars; however, we shall not distinguish them here and label them just as $\vec \phi=(\phi_1, \phi_2, \cdots, \phi_n)$. It follows from \cite{Heidenreich:2020upe} and the discussions in Section 2, the long-range force condition \eqref{lrfl2} now becomes
\be
\Big[M + \Big(\fft{\tilde\delta}{\delta}-1\Big)\mu\Big]^2 +
\fft{4}{\delta}\Big[\sum_{i,j} G_{ij}(\phi)|_\infty \Sigma^i \Sigma^j - \sum_{\alpha=1}^N Q_\alpha^2\Big] =
\Big(\fft{\tilde\delta}{\delta}\mu\Big)^2\,.\label{lrfl3}
\ee
We define the matrix-valued vectors,
\be
{\vec {\cal M}'} = \fft{\partial {\cal M}}{\partial \vec\phi}\Big|_\infty\,.
\ee
The scalar hair equation \eqref{mcr} is generalized to become
\begin{equation}
	\vec{\Sigma}=\frac{1}{2}Q^{\mathrm{T}}\, {\vec {\cal M}'} \frac{\partial M}{\partial Q}+\mu\frac{\partial\vec{\Sigma}}{\partial\mu}\,.\label{mcr2}
\end{equation}
The homogeneity condition, asymptotic-horizon relation and the first law take the same forms as before. Using the same universal boundary condition \eqref{boundary}, we can derive the black $p$-brane thermodynamics without having to construct the corresponding solution. To illustrate this, we consider a concrete example.

\subsection{An $SL(2,\mathbb R)/SO(2)$ example}

In this subsection, we consider a concrete example of $SL(2,\mathbb R)/SO(2)$ scalar coset, coupled to a doublet of $(p+2)$-form field strengths. Such a theory is ubiquitous in supergravities by suitable truncation. The best known example is the doublet of NS-NS and R-R 3-form field strengths in type IIB supergravity, where an $SL(2,\mathbb R)$ multiplet of type IIB supersymmetric string solutions was constructed \cite{Schwarz:1995dk}. The Lagrangian is given by \eqref{genlag2} with the matrix ${\cal M}$
\be
	\mathcal{M}=\mathrm{e}^{\phi_1}
	\begin{pmatrix}
		1&\phi_2\\
		\phi_2&|\lambda|^2
	\end{pmatrix}\,,\qquad
	F=\begin{pmatrix}
		F_1\\
		F_2
	\end{pmatrix},\qquad \lambda(r)=\phi_2(r)+\mathrm{i}\mathrm{e}^{-\phi_1(r)},
\ee
The scalar Lagrangian can also be expressed as ${\cal L} = - \ft12 (\partial\phi_1)^2 -\ft12 (\partial\phi_2)^2 e^{2\phi_1}$. The black $p$-brane has five parameters to start with, namely $(M, Q_1,Q_2)$ as well as the scalar charges $(\Sigma_1,\Sigma_2)$. Defining dimensionless parameters $x_\alpha=Q_\alpha/\mu$ and $(f,g_1,g_2)$ following \eqref{allthermo}, the long-range force law \eqref{lrfl3} becomes
\be
f^2+\frac{4}{\delta}
\left(g_1^2+g_2^2-x_1^2-x_2^2\right)=\left(\frac{\tilde{\delta}}{\delta}\right)^2.
\ee
The scalar hair equation \eqref{mcr1} becomes
\bea
&&x_1\frac{\partial g_1}{\partial x_1}+x_2\frac{\partial g_1}{\partial x_2}=\frac{1}{2}\left(x_1\frac{\partial f}{\partial x_1}-x_2\frac{\partial f}{\partial x_2}\right),\quad
x_1\frac{\partial g_2}{\partial x_1}+x_2\frac{\partial g_2}{\partial x_2}=\frac{1}{2}\left(x_2\frac{\partial f}{\partial x_1}+x_1\frac{\partial f}{\partial x_2}\right).
\eea
Here we have made an assumption that $\phi_{1,2}(\infty)=0$. When $x_1=0$ or $x_2=0$, the system reduces to the single-charge cases of $a=1$ or $a=-1$ respectively. The solutions are then respectively given by
\begin{gather}
g_1(x_1,0)=-\frac{\tilde{\delta}}{2(1+\delta)}+\frac{1}{2}
\sqrt{\frac{4}{1+\delta}x_1^2+\left(\frac{\tilde{\delta}}{1+\delta}\right)^2},\qquad 	g_2(x_1,0)=0\,;\nn\\
g_1(0,x_2)=\frac{\tilde{\delta}}{2(1+\delta)}
-\frac{1}{2}\sqrt{\frac{4}{1+\delta}x_2^2+
\left(\frac{\tilde{\delta}}{1+\delta}\right)^2}\,,\qquad g_2(0,x_2)=0\,.\label{g1g2single}
\end{gather}
We can then use the above as the boundary conditions to solve the general $(x_1,x_2)$ case. Specifically, to solve these equations, we first simplify their forms by applying the polar coordinates, namely
\begin{equation}
	\begin{cases}
		g_1=g_0 \cos\psi,\\
		g_2=g_0 \sin\psi,
	\end{cases}\quad\text{and}\qquad
	\begin{cases}
		x_1=x_0 \cos\theta,\\
		x_2=x_0 \sin\theta.
	\end{cases}
\end{equation}
The original equations are reduced to the following identities
\begin{gather}
f=\frac{\tilde{\delta}}{\delta}+G_1,\qquad \frac{\partial G_1}{\partial \theta}=x_0\frac{\partial G_2}{\partial x_0},\\
	\left(G_1+\frac{\tilde{\delta}}{1+\delta}\right)^2+\frac{\delta}{1+\delta}G_2^2
=\frac{4}{1+\delta}x_0^2+\left(\frac{\tilde{\delta}}{1+\delta}\right)^2,\label{ee}
\end{gather}
where
\begin{equation}
	G_1(x_0,\theta)=2g_0\cos(\psi-2\theta)\,,\qquad\ G_2(x_0,\theta)=2g_0\sin(\psi-2\theta),
\end{equation}
and the universal boundary conditions \eqref{boundary} have been used. Note that the identity (\ref{ee}) is an elliptic equation which implies that the solution can be written as
\begin{equation}
G_1=\sqrt{\frac{4}{1+\delta}x_0^2+\left(\frac{\tilde{\delta}}{1+\delta}\right)^2}
\cos\Psi-\frac{\tilde{\delta}}{1+\delta}\,,\qquad G_2=\sqrt{\frac{4}{\delta}x_0^2+\frac{\tilde{\delta}^2}{\delta(1+\delta)}}\sin\Psi,
\end{equation}
where $\Psi(x_0,\theta)$ satisfies
\begin{equation}
\left(x_0\frac{\partial\Psi}{\partial x_0}\right)\cos\Psi
+\left[\sqrt{\frac{\delta}{1+\delta}}\frac{\partial\Psi}{\partial\theta}
-\frac{4(1+\delta)x_0^2}{4(1+\delta)x_0^2+1}\right]\sin\Psi=0\,.\label{ge}
\end{equation}
We can now use the single-charge cases \eqref{g1g2single} as the boundary conditions. Consequently, it can be shown that $\cos(\Psi(x_0,\theta))=1$ for all $x_0$, under the assumption that $\Psi(x_0,\theta)$ is analytic at $x_0=0$. We therefore obtain $(f,g_1,g_2)$ and hence $(M,\Sigma_1,\Sigma_2)$ as functions of $(Q_1,Q_2,\mu)$. The remaining thermodynamic quantities can also be derived. Thus, we obtain the complete set of thermodynamic quantities without constructing the corresponding black $p$-brane. They are
\bea
M&=&\left(1-\frac{\tilde{\delta}}{1+\delta}\right)\mu+
\sqrt{\frac{4}{1+\delta}(Q_1^2+Q_2^2)+\left(\frac{\tilde{\delta}}{1+\delta}\mu\right)^2},\cr
\Sigma_1&=& \frac{Q_2^2-Q_1^2}{Q_1^2+Q_2^2}\left[\frac{\tilde{\delta}}{2(1+\delta)}
-\frac{1}{2}\sqrt{\frac{4}{1+\delta}(Q_1^2+Q_2^2)
+\left(\frac{\tilde{\delta}}{1+\delta}\mu\right)^2}\right],\cr
\Sigma_2&=&\frac{Q_1Q_2}{Q_1^2+Q_2^2}\left[-\frac{\tilde{\delta}}{1+\delta}
+\sqrt{\frac{4}{1+\delta}(Q_1^2+Q_2^2)+
\left(\frac{\tilde{\delta}}{1+\delta}\mu\right)^2}\right]\,,\cr
S&=&s_0\mu^{\frac{2}{\tilde{\delta}}}\left[\frac{1}{2}
+\sqrt{\frac{1}{4}+\frac{1+\delta}{(\tilde{\delta}\mu)^2}
(Q_1^2+Q_2^2)}\right]^{\frac{2}{\Delta}}\,,\cr
\Phi_{\alpha} &=& \frac{4Q_{\alpha}}{\tilde{\delta}\mu+
\sqrt{(\tilde{\delta}\mu)^2+4(1+\delta)(Q_1^2+Q_2^2)}}\,,\qquad \alpha=1,2.
\eea
When $Q_1=0$ or $Q_2=0$, these reproduce the single-charge black $p$-brane thermodynamics of $a=1$ or $a=-1$ respectively. Note that if we let $Q_0=\sqrt{Q_1^2 + Q_2^2}$ and $\Sigma_0=\sqrt{\Sigma_1^2 + \Sigma_2^2}$, the thermodynamics reduces to the same form as the single-charge black $p$-brane. This indicates that the corresponding black $p$-brane solution can be obtained from the $SO(2)$ global symmetry rotation of the single-charge black $p$-brane. Such a construction for extremal supersymmetric type IIB string solutions was carried out in \cite{Schwarz:1995dk}.

\section{Conclusions}

In this paper, we considered supergravity or supergravity-inspired Lagrangians that admit black $p$-branes charged under single or multiple $(p+2)$-form field strengths. For suitable dilaton couplings, many exact solutions exist \cite{Duff:1996hp,Cvetic:1996gq}, which allows one to derive the corresponding black $p$-brane thermodynamics. However, in general setting, exact solutions are not accessible.

Based on the previous black hole cases, we proposed a set of equations that govern the thermodynamic quantities, which enabled us to derive the black $p$-brane thermodynamic directly without the need of constructing the corresponding black $p$-brane solutions. The generalization from black hole to black $p$-brane is straightforward for most of the thermodynamic equations; however, the long-range force law is subtle. Based on the work of \cite{Heidenreich:2020upe}, we proposed \eqref{lrfl2} and \eqref{lrfl3} for the new long-range force law, which is valid even when solution is not a black extended object. We were then able to derive the black $p$-brane thermodynamics and showed that they match the previously-known results derived from exact black $p$-brane solutions. These results not only confirm our equations of deriving the black $p$-brane thermodynamics, but also verify the long-range force law of \cite{Heidenreich:2020upe}. Our work therefore provides a new simple framework of studying a variety of black $p$-brane thermodynamics in string theories where exact solutions are not accessible.

\section*{Acknowledgement}

This work is supported in part by the National Natural Science Foundation of China (NSFC) grants No.~12375052 and No.~11935009, and also by the Tianjin University Self-Innovation Fund Extreme Basic Research Project Grant No.~2025XJ21-0007.

\end{document}